\documentstyle[subeqnar,rep12]{article}
\newlength{\dinwidth}
\newlength{\dinmargin}
\def\reff#1{(\ref{#1})}
\def\be{\begin{equation}}
\def\ee{\end{equation}}
\def\cvc{\mbox{$CV_N(J_c)$}}
\def\chic{\mbox{$\chi_N(J_c)$}}
\def\mc{\mbox{$M_N(J_c)$}}
\def\cvmax{\mbox{$CV_{max}$}}
\def\chimax{\mbox{$\chi_{max}$}}
\def\jcvmax{\mbox{$J_N(CV_{max})$}}
\def\jchimax{\mbox{$J_N(\chi_{max})$}}
\title{The Ising model on spherical lattices: dimer versus Monte Carlo
approach}
\author{{\bf O. Diego} \thanks{e-mail: imtod67@cc.csic.es}, \
        {\bf J. Gonz\'alez} \thanks{e-mail: imtjg64@cc.csic.es} \\
        {\em Instituto de Estructura de la Materia } \\
        {\em Serrano 123, 28006 Madrid} \\
        {\em Spain } \\
         \mbox{}     \\
         and         \\
         \mbox{}     \\
        {\bf J. Salas} \thanks{e-mail: duncan@vm1.sdi.uam.es} \\
        {\em Departamento de F\'{\i}sica Te\'orica C-XI} \\
        {\em Universidad Aut\'onoma de Madrid} \\
        {\em Cantoblanco 28049 Madrid} \\
        {\em Spain} }

\begin{document}
\setcounter{page}{0}
\maketitle
\thispagestyle{empty}
\begin{abstract}

We study, using dimer and Monte Carlo approaches, the critical
properties and finite size effects of the Ising model on honeycomb
lattices folded on the tetrahedron. We show that the main critical
exponents are not affected by the presence of conical singularities.
The finite size scaling of the position of the maxima
of the specific heat does not
match, however, with the scaling of the correlation length, and the
thermodynamic limit is attained faster on the spherical surface than
in corresponding lattices on the torus.

\end{abstract}

\newpage

\section{Introduction}

The Ising model conveys, in its simplicity, a richness of
physical information which makes it relevant as a model for
critical phenomena in different instances (ferromagnetic materials,
lattice gas, binary alloys).
The model is also paradigmatic of a common situation in
statistical physics since, being one of the simplest models, yet
it only allows to compute analytically the thermodynamic limit for particular
classes of lattices (in one or two dimensions). In two
dimensions, the Ising model in a square lattice has been solved
in the continuum limit with cylindrical and toroidal boundary
conditions \cite{onsa,kast}.
It has been also found analytic solution for the
two-dimensional model on a triangular or honeycomb
lattice \cite{varios}. In
general, however, the introduction of more specific boundary
conditions precludes the resolution of the model in closed
analytic form.

On the other hand, when resorting to a numerical simulation of
the observables one may take advantage of finite size effects to
infer the critical behavior in the thermodynamic limit \cite{barber}.
It happens, though, that finite size effects depend in general on
the boundary conditions, in a way that may not be crucial but
one cannot predict. There
are again a few number of situations in which the asymptotic
dependence on the spatial dimensions of the lattice has been
rigorously studied.  One of these cases corresponds to the
analysis by Ferdinand and Fisher of the two-dimensional Ising
model on large toroidal lattices \cite{fisher}.
The conclusions reached there
support, in essence, the assumptions made in discussing finite
size effects and, more precisely, the hypothesis of finite size
scaling \cite{fss}. Some open questions are raised, however, regarding
the approach to the critical coupling, which is drastically
influenced by the shape of the torus.

The purpose of the present paper is to investigate finite size
effects and  critical
properties of the Ising model on a class of two-dimensional
lattices with spherical topology. Our choice for the elements of
this class is not arbitrary, but is rather dictated by a
prescription which makes possible increasing the size of the
lattice without changing the local geometry. We propose to
consider, in fact, a type of honeycomb lattices folded on the
tetrahedron, which are built by assembling triangular blocks of
the kind shown in Figure~1 as the faces of the polyhedron. One
may construct a whole family of these lattices with increasing
size, in such a way that the member of the $N^{\rm th}$ generation
$\Delta_n$
has a number of lattice points equal to $n = 12 N^{2}$. The
coordination number is constant in each lattice. Moreover, from
the point of view of the simplicial geometry, the curvature is
always concentrated at the same faces, which are those formed by
the three-fold rings around the four vertices.
In principle, this makes the problem
of taking the thermodynamic limit along our sequence of lattices
well-defined. In ref. \cite{jose} clear evidence was given of
critical behavior in the ferromagnetic regime, as well as
evidence supporting the hypothesis of finite size scaling
applied to the model on the curved surface.

The lattices we are considering may be understood as the result
of applying nontrivial boundary conditions for the honeycomb
lattice on the plane, though they have the effect of introducing
curvature in the model. We undertake the
investigation of the influence of these boundary conditions on
finite size effects and, more significantly from the physical
point of view, on the critical properties of the model.
Regarding the first point  we will see that a discrepancy
arises between the scaling of {\em pseudocritical} coupling
constants for finite lattices and the true scaling of the
correlation length. The second issue may be probably addressed
in the continuum. In fact, investigations of the effect of
boundary conditions in conformal field theories have been
carried out before \cite{cardy1}. Anyhow, the inclusion of a conical
singularity requires the kind of boundary condition which may
call for a nonlocal operator in the theory \cite{pol}, so the analysis
of our model in the continuum does not appear quite straightforward.

The content of the paper is distributed as follows.
In Section 2 we review the
dimer approach applied to the computation of the partition
functions and correlation functions of the model. Section 3 is
devoted to the finite size analysis of data obtained with the
above method, clarifying the issue concerning the $\nu$ critical
exponent. In Section 4 we give technical details of the Monte
Carlo simulations carried out to measure in some of the larger
lattices. Section 5 contains the results for the critical
exponents $\alpha, \beta, \gamma$ obtained after combining data
from the dimer approach and the Monte Carlo simulations. Finally
we draw our conclusions and outline further directions of our work.

\section{The dimer approach to the Ising model}

We review in this section the dimer formulation of the
two-dimensional Ising model \cite{mont,mcoy-wu}.  This approach
presents the advantage of being applicable to lattices with an
irregular coordination. It allows to write partition functions and
correlation functions in closed compact form, in terms
essentially of determinants of some coordination matrices for
the lattice. This is something that cannot be achieved for our
curved lattices by any other standard resolution method of the
two-dimensional Ising model. There is no obvious way, for
instance, as to how to apply the transfer matrix method to write
down the partition function for a lattice with the topology of the
sphere, not even to produce a numerical computation
of the same. Within the dimer approach one may, in principle,
compute the partition function for any of the hexagonal lattices
inscribed on the tetrahedron. Although we have not been able to
infer from this construction the thermodynamic limit along the
sequence of growing lattices,  the method shows very efficient
to calculate observables like the specific  heat or the
correlation length with arbitrary precision. One can easily
progress to lattices with up to more than 1000 points, with the
possibility of applying a finite size analysis to study the
critical behavior of the model.

The dimer formulation of the Ising model makes use first of the
equivalence between the partition function of the model and the
dimer partition function of certain decorated lattice built from
the original one \cite{kast}. One may apply afterward powerful
techniques
developed to perform the sum over dimer configurations. Let us
review the former correspondence for the hexagonal lattice,
while allowing for some kind of frustration which keeps constant
the coordination number over the lattice. Given the collection
of spins $\{\sigma_{i} \}$, with $i$ running over all the
lattice points, the partition function ${\cal Z}$ is defined as
the sum over all possible configurations
\begin{equation}
  {\cal Z} = \sum_{\sigma_{i} = \pm 1} \mbox{\large $e^{- J H}$}
  \label{1}
\end{equation}
where the Hamiltonian $H$ is given by the sum over all the
lattice links $\langle i,j \rangle$
\begin{equation}
  H = - \sum_{<i,j>} \sigma_{i} \sigma_{j}
\end{equation}
The factor $1/kT$ is absorbed for simplicity in the definition of the
nearest--neighbor coupling $J$.
It is well known that (\ref{1}) can be cast as a sum over all
the closed loops over the  lattice. Calling this collection  $\{
l_{i} \}$ and $\{ n_{i} \}$ the respective numbers of links of
the paths, we have actually
\begin{equation}
  {\cal Z} = (\cosh J)^l \sum_{ \{ l_{i} \} } (\tanh J)^{ n_i }
  \label{2}
\end{equation}
$l$ being the total number of links of the lattice. One can
draw a correspondence between each closed path and a dimer
configuration in the appropriate decorated lattice. This is
formed in our model by inserting a triangle in place of  each
of the points of the original lattice. To each of these one may
assign four different states, depending on whether the point is
traversed or not by a closed path and on what direction, in the
first instance. These states are labeled in Figure 2. In a
similar fashion, there are four different possible
configurations of dimers on each triangle and adjacent links of
the decorated  lattice, which bear a one-to-one correspondence
with the above states. These dimer configurations are labeled in
Figure 3. One may easily convince that, establishing as a rule
the  equivalence between respective states in Figures 2 and  3,
a unique closed path can  be reconstructed starting from a
given dimer configuration  in the  decorated  lattice, and
viceversa. Furthermore, if a weight equal to $ z = \tanh J $
is given to each dimer on a triangle link and equal to 1
for dimers joining
neighboring triangles, it is clear that the dimer partition
function reproduces  the sum in the Ising partition  function
(\ref{2}).

There exist, in turn, powerful techniques developed to the
computation of dimer partition functions, which rely mainly on
the relation between these and the Pfaffians of appropriate
coordination matrices on the decorated lattice. We sketch here
this relation, which has been worked out quite rigorously in
ref. \cite{mcoy-wu}. The first step is to establish an order relation
among the points of the decorated lattice. Once this is done,
one may assign a matrix element $a_{p_{1}p_{2}}$ for points
numbered $p_{1}$ and  $p_{2}$ such that $a_{p_{1}p_{2}} = \tanh J$ if
the points belong to the same triangle, $a_{p_{1}p_{2}} = 1$ if
they are nearest neighbors belonging to different triangles, and
$a_{p_{1}p_{2}} = 0$ otherwise. The sum over all dimer
configurations weighted as  proposed before amounts to perform
the sum over all permutations $\{p_{1}, p_{2}, \ldots p_{K} \}$
\footnote{$K$ is the total number of points in the decorated lattice.}
\begin{equation}
  \sum a_{p_{1}p_{2}} a_{p_{3}p_{4}} \ldots a_{p_{K-1}p_{K}}
\end{equation}
restricted by $p_{1} < p_{3} < \ldots p_{K-1}$ and
$p_{1} < p_{2}, p_{3} < p_{4}, \ldots p_{K-1} < p_{K}$.
While there is no known algorithm to compute efficiently a sum
of the above kind, one may think of allowing for an
antisymmetric matrix $A = \{ a_{ij} \}$, so that the dimer partition
function (a sum of all positive terms) may become proportional
to
\begin{equation}
  \sum (-1)^{P} a_{p_{1}p_{2}} a_{p_{3}p_{4}} \ldots a_{p_{K-1}p_{K}}
  \label{7}
\end{equation}
with $p_{1} < p_{3} < \ldots p_{K-1}$,
$p_{1} < p_{2}, p_{3} < p_{4}, \ldots p_{K-1} < p_{K}$,  as
before, and $(-1)^{P}$ being the signature of the permutation.
The expression (\ref{7}) reproduces the definition of the
Pfaffian of the matrix $A$, which may be subsequently
computed as the square root of its determinant.
The remarkable conclusion which follows from the work of ref.
\cite{kast} is that, for planar lattices, it is always possible
to choose the sign of the nearest-neighbor matrix elements
$a_{p_{1}p_{2}}$ so that all the  terms in the sum (\ref{7})
have the same sign. Since the matrix $A$ becomes
antisymmetric, it is customary to fix pictorically the sign of
each $a_{p_{1}p_{2}}$ by giving an orientation to every link of the
decorated lattice ---$a_{p_{1}p_{2}}$ is positive, for instance,
if the arrow goes from $p_{1}$ to $p_{2}$---. We may enunciate the
Kasteleyn theorem by saying that in any planar lattice there is
always a system of arrows such that the dimer partition function
can be computed as the Pfaffian of the corresponding
antisymmetric coordination matrix.

The lattices we consider here fall into the category of planar
lattices since they have the topology of the sphere. As long as
we  are interested on dimers mainly for computational purposes,
we simply give the recipes which have to be followed to form
the appropriate system of arrows on a planar lattice. Once
superposed on the  plane,  the lattice is made of so-called
elementary polygons, which are the closed cycles that do not
contain points in  their interior. On the other hand, a polygon
is said to be clockwise odd if the number of arrows pointing in
the clockwise direction in the polygon is odd. The basic results
which hold for planar lattices are that (a) it is  always
possible to choose a system of arrows such that all the
elementary polygons are clockwise odd, and (b) with this choice
and taking $a_{p_{1}p_{2}}$ as positive when the arrow goes
from $p_{1}$ to $p_{2}$,
all the terms in the expansion of the Pfaffian (\ref{7}) have
the same sign.

In our case, a possible system of arrows realizing the property
(a) for a decorated lattice inscribed on the tetrahedron is
shown in Figure 4, where all the arrows for the triangle links
(not drawn) are supposed to point in the clockwise direction.
The advantage of this choice of arrows is
that it keeps a regular pattern in the bulk, while only a few
arrows on the boundary links have to be flipped to make all
elementary polygons clockwise odd. In general, progressing to
the next member of our family of lattices on the tetrahedron
just amounts to adding a column of (decorated) hexagons at each side of
Figure 4, expanding appropriately the vertical dimension.
The system of arrows proposed may be extended in a
straightforward way to larger lattices.
According to the above discussion, we set always the absolute value of
matrix elements $a_{p_{1}p_{2}}$ equal to $\tanh J$ for
points in the same triangle, and equal to 1 for
nearest neighbors on different triangles. The partition function
for any honeycomb lattice on the tetrahedron can be represented,
therefore, in terms of the respective matrix $A$ by
\begin{equation}
  {\cal Z} = (\cosh J)^{l} (\det A)^{1/2}
  \label{12}
\end{equation}

We have made use of the representation (\ref{12}) to perform the
numerical computation of the maximum of the specific heat (in
the ferromagnetic regime) for lattices up to $\Delta_{1452}$. We
have been able to measure that quantity with a relative error of
less than $10^{-7}$ in most of the cases. Correspondingly, a
precise determination of the coupling constant at which the maximum is
attained in each lattice has been also possible (see Table 1).
The values of
these pseudocritical coupling constants are fundamental
ingredients for the finite size analysis to be accomplished in
the next section. We have also computed the values of the
specific heat of the curved lattices at the critical coupling
constant of the planar honeycomb lattice (see Table 2). These are also
relevant under the hypothesis of finite size scaling since, as
we will see, the sequence of pseudocritical temperatures
converges in the thermodynamic limit to the critical temperature
of the planar hexagonal lattice.

We conclude this section with an outline of how the two-point
correlation functions can be obtained within the dimer
approach \cite{mont,mcoy-wu}.
Given two arbitrary spins $\sigma_{p}$ and $\sigma_{q}$ in the
lattice, the average
\begin{equation}
  \left\langle  \sigma_p \sigma_q \right\rangle =
  \frac{1}{\cal Z}
  \sum_{\sigma_i = \pm 1} \sigma_p \sigma_q  e^{- J H}
\end{equation}
may be computed with the following trick. One chooses a path
${\cal C}$ from $\sigma_{p}$ to $\sigma_q$ on the lattice,
which will comprise a number of consecutive spins $\{
\sigma_{p_{1}}, \sigma_{p_{2}}, \ldots \sigma_{p_m} \}$. The
two-point function may also be expressed as
\begin{equation}
  \left\langle  \sigma_p \sigma_q \right\rangle =
  {1 \over {\cal Z} }
  \sum_{\sigma_i = \pm 1} \sigma_p \sigma_{p_1}
  \sigma_{p_1} \ldots \sigma_{p_{m-1}} \sigma_{p_m}
  \sigma_q  e^{- J H}
\end{equation}
Now we have that $\sigma_p \sigma_{p_1}$, $\sigma_{p_1}
\sigma_{p_2}$, \ldots $\sigma_{p_m} \sigma_q$ are pairs of
nearest-neighbor spins. Therefore, we find
\begin{equation}
  \left\langle  \sigma_p \sigma_q \right\rangle =
  {1 \over {\cal Z} }
  \sum_{\sigma_{i} = \pm 1}
  \prod_{<i,j> \not\in {\cal C}} (\cosh J + \sigma_i \sigma_j
                                  \sinh J)
  \prod_{<k,l> \in {\cal C}}  (\sinh J + \sigma_k
                      \sigma_l \cosh J)
  \label{17}
\end{equation}
where the first product extends to all the links which do not
belong to ${\cal C}$, and the second product runs over the links
that do belong to the path.
{}From (\ref{17}) we arrive finally at the expression
\begin{equation}
  \left\langle  \sigma_p \sigma_q \right\rangle =
  {1 \over {\cal Z} }
  (\cosh J)^{l} (\tanh J)^{m+1}
  \sum_{\{ l_i \} }    (\tanh J)^{n_i - r_i} \frac{1}{(\tanh J)^{r_i}}
  \label{18}
\end{equation}
where the sum, as in expression (\ref{2}), is over all the
closed loops on the lattice, but with the difference now that
the number $r_{i}$ of links in each loop belonging to ${\cal C}$
have to be weighted with $(\tanh J)^{-1}$ rather than with
$\tanh J$. It becomes obvious that all the machinery
of the dimer formulation can be applied again to transform
the sum in (\ref{18}) into a suitable dimer partition function
on the decorated lattice, so that
\begin{equation}
  \left\langle  \sigma_p \sigma_q \right\rangle =
  {1 \over {\cal Z} }
  (\cosh J)^{l} (\tanh J)^{m+1}   (\det A^\prime)^{1/2}
  \label{19}
\end{equation}
The appropriate coordination matrix $A^\prime = \{ a_{ij}^\prime \}$ has
to keep track of the different weight that the $m+1$ links in ${\cal C}$
carry in the sum over closed loops.

\section{Finite size scaling and critical exponents}

\subsection{Finite size scaling}

It is well known that singularities in the free energy (i.e. phase
transitions) can only occur in the thermodynamic limit. For finite
volumes the free energy is an analytic function of the temperature and
any other parameter in the Hamiltonian. The thermodynamic singularities
are thus smoothed out around the transition point. A trace of the
existence of such non-analyticities is the presence of some peaks in the
specific heat $CV$ or magnetic susceptibility $\chi$ curves.
The dependence on the linear size of the system $L$ of the location of
the maxima of those peaks and their height permits the description of
the thermodynamic limit from finite--size data \cite{barber}.

In second order phase transitions this round--off is due to the fact
that the correlation length $\xi$ is limited by the size of the system.
This fact defines a pseudocritical coupling $J^\star(L)$ such that
\be
  \label{def_j_star}
  \xi(J^\star(L)) \sim L
\ee
At this point the surface contribution to the free energy is not
negligible compared to the bulk one. In the vicinity of
a second order transition point $J_c$ the correlation length diverges
with a power--law given by
\be
  \label{def_nu}
  \xi(J) \sim (J - J_c)^{-\nu}
\ee
{}From \reff{def_j_star} and \reff{def_nu} it can be derived
the dependence of the pseudocritical coupling on the lattice size
\be
  \label{scaling_j_star}
  | J^\star(L) - J_c | \sim L^{-1/\nu}
\ee

Unfortunately in practical situations it is a very involved task
to compute such quantity $J^\star(L)$. It is easier to look
at the position and height of those peaks mentioned above.
If some quantity $P$ diverges near the critical point as
\footnote{Hereafter we will denote quantities computed in a finite
volume with a subscript $L$ meaning the linear size of the system
(i.e. $P_L$). Whenever no subscript is present, the thermodynamic
limit is assumed (i.e. $P = \lim_{L\rightarrow \infty} P_L$).}
\be
  \label{def_rho}
   P(J) \sim | J -  J_c|^{-\rho} ; \qquad \rho > 0
\ee
then it can be shown \cite{barber} that for a finite volume it attains
a maximum value $P_{max}(J_L)$ at a point $J_L(P_{max})$ given by
\begin{subeqnarray}
  \slabel{scaling_j_p}
  |J_L(P_{max}) - J_c| & \sim & L^{-\theta_P} \\
  \slabel{scaling_p_max}
  P_{max}(J_L)         & \sim & L^{\rho/\nu}
\end{subeqnarray}
when $L$ is large enough. In most systems it is found that
\be
  \label{igualdad}
  \theta_P = {1 \over \nu}
\ee
but this is not a general result. There are some examples where this
property does not hold: the spherical model, the ideal Bose gas
\cite{modelos_raros} and the one--dimensional $q = \infty$ clock model
\cite{clock}. In the present paper we will face another situation in which
the relation (\ref{igualdad}) is violated.
On the other hand, if $\theta \geq 1/\nu$ then the behavior of
this quantity at finite volume evaluated at the critical point
$P_L(J_c)$ is the same as in \reff{scaling_p_max}
\be
  \label{scaling_p_c}
  P_L(J_c) \sim L^{\rho/\nu}
\ee
Using \reff{scaling_j_p} and \reff{scaling_p_max} the critical
coupling $J_c$ and the critical exponents ratio $\rho/\nu$ can be
derived from finite--size data. When $J_c$ is explicitly known and
$\theta_P \geq 1/\nu$ then \reff{scaling_p_c} can be used
instead of \reff{scaling_p_max}. In this paper we are mainly
concerned with the analysis of the susceptibility and the specific
heat. So, we will obtain estimates of $J_c$ and the
ratios $\gamma/\nu$ and $\alpha/\nu$.
The rest of the critical
exponents may be derived using the scaling relations \cite{itzykson}:
\begin{subeqnarray}
   \slabel{beta}
   {\beta \over \nu} &=& 1 - {\gamma \over 2\nu} \\
   \slabel{one_nu}
   {1 \over \nu}     &=& 1 + {\alpha \over 2\nu} \\
   \slabel{eta}
   \eta       &=& 2 - {\gamma \over \nu} \\
   \slabel{delta}
   \delta      &=& {4 \over \eta} - 1
\end{subeqnarray}
However, in this paper we will check numerically scaling relations
\reff{beta} and \reff{one_nu}.
We will obtain independent estimates of $\nu$
and $\beta/\nu$ respectively in terms of the analysis of the
correlation
length (see below) and the magnetization at the critical point (see
Section~5).

\subsection{Critical point and exponent $\theta_{CV}$}

In Section 2 we showed that for lattices up to $\Delta_{1452}$
we were able to obtain very accurate estimates of the internal
energy E and the specific heat. These quantities are defined
hereafter as follows
\begin{eqnarray}
  \label{energia}
    E_N &=&  {2 \over 3V} \left\langle  H \right\rangle \\
  \label{calor_e}
 CV_N   &=& {3V \over 2} \sigma^2 \left( {2 \over 3V } H  \right)
\end{eqnarray}
where the factor $3V/2$ is equal to the number of links in a lattice of
$V$ sites and $\sigma(\cdot)$ is the standard deviation.
Thus, the values of $CV_{max}$ and $J_L(CV_{max})$ can be computed with
high precision. In what
follows we will  identify the lattice linear size $L$ with the index
$N$ characterizing the fullerene lattice (see Section 1). This choice is
consistent as the volume increases as $V = 12\  N^2$.

Data will be fitted according to Equation~\reff{scaling_j_p}
\be
   \label{fit_j_c}
   J_N(CV_{max}) = J_c + A N^{-\theta_{CV}}
\ee
using a least $\chi^2$ method. Here the input errors are given by the
precision of the computer in calculating $J_N(CV_{max})$. In order to
obtain a more reliable
result, we will sequentially remove the point with smallest $N$.
One eventually can observe
a monotonous trend to some value, which will be identified with the
thermodynamic limit.

Our best result is
\be
    J_c = 0.65850 \pm 0.00002
  \ee
for $5 \leq N \leq 11$. In this case $\chi^2 = 0.07$ with 2 degrees of
freedom.
Throughout this paper all the errors associated with our final
results will be
equal to 2 standard deviations ---i.e.~95\% of confidence level---.
The later result is compatible with the critical point of the
Ising model on a toroidal honeycomb lattice
\be
  \label{j_c_toro}
  J_c^{\rm torus} = {1 \over 2} \log (2 + \sqrt{3}) = 0.65848
\ee
This result can be easily derived using duality \cite{itzykson,baxter}.
Thus, our data strongly supports that both critical points coincide.

A good estimate of $\theta_{CV}$ is obtained by repeating the fit with
$J_c = J_c^{\rm torus}$. The result is
\be
        \theta_{CV} = 1.745  \pm 0.015
\ee
for $6 \leq N \leq 11$ and with $\chi^2 = 1.2$.
This one is in clear disagreement with the  result expected for a
lattice on a torus ($\theta_{CV} = 1$ \cite{fisher}). This fact makes
necessary a direct determination of the $\nu$ critical exponent, in
order to see if the above measurement bears any relation to it (see
below).

We will skip here the analysis  of $CV_{max}$, as it is rather subtle
to distinguish between a logarithmic and a power--law behaviors,
specially when such a power is rather small.
This will carried out in Section 5.

\subsection{Correlation length and exponent $\nu$}

An independent way to compute the  critical exponent $\nu$ is to
study the correlation length near the critical point. This quantity is
defined in terms of the connected two--point function
\be
  \label{correlation_length}
  \left\langle \sigma_{\bf 0} \cdot \sigma_{\bf r} \right\rangle^c =
  \left\langle \sigma_{\bf 0} \cdot \sigma_{\bf r} \right\rangle -
  \left\langle \sigma_{\bf 0} \right\rangle
  \left\langle \sigma_{\bf r} \right\rangle
  \sim e^{-r/\xi}
\ee
when $r$ is large enough. The connected two--point correlators are
equal to the usual ones  $\left\langle \sigma_{\bf 0} \cdot
\sigma_{\bf r} \right\rangle$ for $J < J_c$ (unbroken phase).
This feature allows
their exact computation using the machinery developed in Section 2.
(For finite lattices odd quantities such as the magnetization are
always equal to zero, even in the broken phase).

A major problem is how to recover $\xi_N(J)$ from the
finite--volume correlators $\left\langle \sigma_0 \cdot \sigma_{\bf r}
\right\rangle_N$. For a torus of linear size $L$ these functions are
expected to behave as $\sim \cosh((x-L/2)/\xi_L)$
when $x \gg 1$. But this question is not clear for
the truncated tetrahedron. On the other hand, it is well known
\cite{mcoy-wu} that the  correlation length does depend on the
direction along the spins $\sigma_{\bf r}$ are disposed. However, the
same critical behavior is expected for all the possible directions.

This study has been  carried out on the lattice $\Delta_{972}$, which
is the largest one allowed by our computer facilities. We believe that
this one is large enough to see the thermodynamic limit.
We have chosen couples of spins
along the diagonal in the representation on Figure 4 of the
tetrahedron unfolded on the plane.
Our choice for
$\sigma_{\bf r}$  allowed us to introduce an increasing distance
between spins running from 3,6,$\cdots$ up to 27 (in units of the
lattice spacing). At
$r = 27$ both spins are located at antipodal points. If we go on along
the diagonal we finally  arrive at $\sigma_0$. For that reason we
expect that the correlators behave for large $r$ as a symmetrized
version of equation \reff{correlation_length}.
In order to improve our results,
we include in our ansatz the correct leading term for
the square lattice on the torus \cite{mcoy-wu}
\be
  \left\langle  \sigma_{\bf 0} \sigma_{\bf r} \right\rangle =
  \frac{f}{\sqrt{r}}  e^{-r/\xi} (1 + {\cal O}(1/r) )
  \label{ansatz_correlation}
\ee
suitably symmetrized around $r = 27$.

We have analyzed the cases $J$ = 0.58, 0.59, 0.60, 0.61 and 0.62
(see Table~3 and Figure~5). We have obtained extremely good fits
for all these cases,
giving differences of order $\sim 10^{-5}$. Although the lattice
$\Delta_{972}$, as any of the lattices inscribed on the
tetrahedron, is not homogeneous, it is remarkable that the
values of the two-point functions at each different $J$ fit,
to a high degree of precision, to the correct leading
behavior for the Ising model on a square lattice on the torus.
The deviation that we have found from the dependence
(\ref{ansatz_correlation}) appears to
be even smaller than for similar measurements carried out for
the lattice on a torus.
The estimated values of the correlation
length are shown in Table~3 and will be used
in the computation of the $\nu$ critical exponent.
We unsuccessfully tried to fit data for $J >
0.62$ to \reff{ansatz_correlation}. The reason is that very close to
the critical point we have to take into  account in Equation
\reff{ansatz_correlation} the ${\cal O}(1/r)$ (or even higher) terms.

Given the values of $\xi^{-1}_{N=9}(J)$ in Table~3, we tried to fit
them according to \reff{def_nu}
\be
     \xi^{-1}_{N=9}(J) = A | J - J_c |^\nu
\ee
We obtain a value equal to $\nu = 1.01 \pm 0.04$ with
$\chi^2 = 0.36$. However, if we drop the point with
$J=0.62$ (the closest to $J_c$) the result is
\be
      \nu = 1.00 \pm 0.06
\ee
with a remarkable small value of the $\chi^2 \sim 7 \cdot 10^{-5}$.

The conclusions of the analysis of the data coming from the dimer
computations can be summarized in the following points (a) The critical
point is compatible with $J_c^{\rm torus}$. (b) The critical exponent
$\nu = 1.00 \pm 0.06$. (c) The finite--size exponent
$\theta_{CV} = 1.745 \pm 0.015$ is significantly different from $1/\nu$.
In summary, our results suggest that the critical properties of the
Ising
model on the truncated tetrahedron are the same as on the torus.
However, we find a very clear difference in the finite--size behavior
of those  models, as long as the scaling of pseudocritical coupling
constants (determined from the maxima of the specific heat)
does not match with the scaling behavior of the correlation
length. On the tetrahedron the thermodynamic limit is
achieved much faster than on the
torus, at least in what concerns to the specific heat.

\section{Technical aspects of the Monte Carlo simulations}

We have performed several Monte Carlo (MC) runs for different lattice
volumes $V = 12 N^2$ and coupling constants $J$. The relevant
information about the simulations can be found in Table~4.

We have used a Metropolis algorithm with the R250
pseudorandom--number generator \cite{random} --initialized with the
RANMAR subroutine--. The period of such generator is equal to
$2^{250}-1$.
Recently, it has been claimed \cite{random} that the combination of the
Metropolis algorithm with the R250 generator gives better results than
other more involved procedures. We have compared those values obtained
both by the dimer approach and by direct MC simulation. They were
consistent within statistical errors.

In all cases we have measured the internal energy density and the
magnetization defined as
\be
   \label{magnetizacion}
    M_N = \left\langle \left| {1 \over V } \sum_i \sigma_i
                       \right| \right\rangle
\ee
We have also measured the specific heat and the magnetic
pure--phase susceptibility,
\be
  \label{suscep}
  \chi_N = V \sigma^2 \left( \left| {1 \over V } \sum_i \sigma_i
                             \right| \right)
\ee

In all cases we discarded the first $10^5$ MC steps for termalization.
Then we have measured each observable once every typically 100
MC steps. In this way we obtained statistically independent data, as
it can be checked by computing the corresponding autocorrelation times
\cite{autocor}.

In order to calculate de maximum value of the specific heat and the
susceptibility we have used the Spectral Density Method
\cite{sweferr}. At a given coupling $J$ we can obtain the histograms
$N(E,M;J)$ which keep track of the numbers of configurations with
magnetization $M$ and internal density energy $E$.  This information
is enough in order to compute the expectation value of any function
$f(M,E)$ at any other coupling $J^\prime$. In our case the magnetic
field is zero and in the equations
(\ref{energia},\ref{calor_e},\ref{magnetizacion},\ref{suscep})
the observables do not depend on
$E$ and $M$ simultaneously. For those reasons we could use the following
formulae
\begin{eqnarray}
   \langle f(E) \rangle (J^\prime) &=& \frac{\sum_{E} f(E) N_{0}(E;J)
   \exp\{(J^\prime - J)E \} }{ \sum_{E} N_{0}(E;J)
   \exp\{(J^\prime - J)E \} } \\
   \langle f(M) \rangle ( J^\prime ) &=& \frac{\sum_{E} N_{f}(E;J)
   \exp\{(J^\prime - J)E \} }{\sum_{E} N_{0}(E;J)
   \exp\{(J^\prime - J)E \}}
\end{eqnarray}
where the one dimensional histograms are defined as follows
\begin{eqnarray}
   N_{0} (E;J) &=& \sum_{M} N(E,M;J) \\
   N_{f} (E;J) &=& \sum_{M} f(M) N(E,M;J)
\end{eqnarray}

The Spectral Density Method gives the correct answer for couplings
$J^\prime$ close to the coupling $J$ where the simulation was performed.
A criterion for the applicability of such method is the following
\cite{sweferr}
\be
       \label{criterio}
       |J^\prime - J| \sim {1 \over \sigma(E) V}
\ee
In most cases only one simulation at $J=J_c$ is enough in order to
determine
the maximum of $CV$ and $\chi$. However, for the smaller lattices an
additional run had to be performed in order to obtain a reliable
estimate of such quantities.

We have divided up the entire sample into typically 30--120 subsamples,
each of them containing $\sim 1000$ measures. For each subsample we
computed
every quantity (including $CV_{max}$, $J_N(CV_{max})$, $\ldots$). With
these estimates we calculated the statistical errors using the
jack--knife method \cite{okawa}. In this way, the effect of
correlation among data was taken into account.

We have done all the MC simulations on a VAX 9000 machine with a
vectorial processor. The program was not fully vectorizable, as the
lattice could not be splitted into two disjoint sublattices, in such a
way every element of one sublattice is surrounded by elements belonging
to the other one\footnote{This feature has to do with the
onset of frustration in the antiferromagnetic regime.}.
However, we could divide the whole lattice into
three subsets. The elements of the first two are arranged on two
disjoint triangular sublattices, so their update could be
fully vectorize.
On the other hand, the rest of the spins can be located on two lines
joining pairs of vertices on the tetrahedron
and their number depend explicitly on the planar representation of the
lattice. For these ones the update is clearly not vectorizable. However,
their
effect on the CPU time is not very important for the larger lattices, as
their number behaves as $\sim V^{1/2}$.

\section{Results of the Monte Carlo simulations}

\subsection{Position of the critical point}

Here we will repeat the analysis of Section 3, but with all the data of
Table 1. For data coming from the dimer analysis the input error will be
taken
as the precision of the subroutines used. For those coming from the MC
simulations  the error is given by the jack--knife method described in
the preceding section. Data for
$N > 11$ posses large error bars compared with
the rest. For that reason the fits presented in Section 3 cannot have
large variations. Our final results for the specific heat are
\be
       J_c = 0.65850 \pm 0.00002
\ee
for $5 \leq N \leq 21$ and with $\chi^2 = 0.67$. And
\be
      \theta_{CV} = 1.745 \pm 0.015
\ee
for $6 \leq N \leq 21$ and with $\chi^2 = 1.9$ (See Figure~7).

If we repeat the same procedure with the susceptibility we
obtain an estimate for $J_c$ compatible with the later one,
but with larger error bar. If we fix this quantity to
$J_c^{\rm torus}$ we arrive at the following estimate for
the exponent $\theta_\chi$
\be
 \theta_\chi = 1.01 \pm 0.02
\ee
for $9 \leq N \leq 21$ and $\chi^2 = 0.8$ (See Figure~7).

We observe that $\theta_\chi$ is close the value $1/\nu = 1$, in
agreement with the Ising model on the torus. We conclude that
\be
   {1 \over \nu} = \theta_\chi < \theta_{CV}
\ee
and this feature implies that one can obtain the critical exponents
using either eq.~\reff{scaling_p_max} or eq.~\reff{scaling_p_c}. As we
have
identified the critical coupling of our system, it seems more natural
to
rely our conclusions on eq.~\reff{scaling_p_c}. In any case, the values
obtained from eq.~\reff{scaling_p_max} are always consistent with those
presented in this paper within statistical errors.

\subsection{Exponent ratios $\gamma/\nu$ and $\beta/\nu$}

The value of $\gamma/\nu$ can be derived using the value of the
critical magnetic susceptibility $\chi_N(J_c)$. We have fitted our data
to
\be
   \chi_N(J_c) =  A N^{\gamma/\nu}
\ee
and our best result is
\be
  {\gamma \over \nu} = 1.73 \pm 0.02
\ee
for values of $N$ ranging from 21 to 9 and with $\chi^2 = 0.79$
This result is very close to the usual Ising model one
$\gamma/\nu = 7/4 = 1.75$. If we fix $\gamma/\nu$ to this value we
obtain a $\chi^2$ value of $\sim 2.8$ which shows that the fit is
reasonably good (See Figure~8).

To determine the value of $\beta/\nu$ in an independent way (i.e. not
using scaling relation \reff{beta}) we will fix our attention on the
magnetization at the critical point. In the thermodynamic limit
the magnetization near $J_c$ behaves as
\be
   \label{def_beta}
   M(J) \sim |J - J_c|^\beta
\ee
Using similar arguments as in Section~3 it can be predicted that
\be
   \label{scaling_mc}
   M_N(J_c) \sim N^{- \beta/\nu}
\ee
The result of performing such fit is the following
\be
   {\beta \over \nu} = 0.126 \pm 0.004
\ee
using data with $7 \leq N \leq 21$ and with $\chi^2 = 0.33$ (See
Figure~9). This
value is also very close to the Ising one $\beta/\nu = 1/8 = 0.125$.
This result supports that the scaling relation \reff{beta}
does hold in this model.

This feature can be used in order to obtain a more accurate estimate of
those exponents. We can try to fit $\chi_N(J_c)$ and $M_N(J_c)$
simultaneously using explicitly the relation \reff{beta}. The result is
\be
    \begin{array}{l}
     \gamma/ \nu  = 1.748 \pm 0.008  \\
     \beta / \nu  = 0.126 \pm 0.004
    \end{array}
\ee

Thus, our results strongly suggest that the ratio $\gamma/\nu = 7/4$ as
in the Ising model on a torus. Notice that the error in that ratio is
less than 0.6\%.
Using the relations \reff{eta} and
\reff{delta} we can derive the value of the exponents $\eta$ and
$\delta$.
\begin{subeqnarray}
\eta   &=& 2 - {\gamma \over \nu} = 0.252 \pm 0.008 \\
\delta &=& {4 \over \eta} - 1 =  14.8 \pm 0.6
\end{subeqnarray}

\subsection{Exponent ratio $\alpha/\nu$}

We can perform the same game for the specific heat and try to obtain the
exponent ratio $\alpha/\nu$. If we try to fit data to the function
\be
  \label{fit_cv}
  CV_N(J_c) = A + B N^{\alpha/\nu}
\ee
we do not obtain a satisfactory result. The best one gives a ratio
$\alpha/\nu \sim 0.060$ with $\chi^2 \sim 9$ and $9 \leq N \leq 21$
(See Figure~8).
On the other hand, and motivated in part for the preceding results we
could try to fit the data to a logarithmic function.
\be
  CV_N(J_c) = A + B \log N
\ee
In this case, the
fit is successful giving a $\chi^2 = 1.9$ with $7 \leq N \leq 21$ (See
Figure~10). This immediately implies that
\be
      \alpha = 0
\ee
and using equation \reff{one_nu}
\be
        \nu = 1
\ee
which is in agreement with the result from direct measurements of the
correlation length displayed in Section 3.
Thus, both exponents take the same values as in the
Ising model on the torus.
On the other hand, we have also verified that the scaling
relation \reff{one_nu} does hold in this model.

\section{Conclusions and outlook}

In this paper we have presented the first deep study of the
critical properties of an Ising model on a lattice with the
topology of the sphere. In particular we have chosen the family
of honeycomb lattices that can be constructed on the tetrahedron.
Our results can be summarized as follows
\begin{itemize}

\item
The dimer approach is very useful and competitive for lattices up to
$\sim 10^3$ points. It provides very accurate data for the internal
energy, specific heat and two--point correlators.

\item
The critical properties  of the Ising model on the tetrahedron are just
the same as on the torus.  In particular we have checked that $J_c$ is
the same, as well as the critical exponents $\nu$, $\alpha$, $\gamma$
and $\beta$.

\item
We also have checked two scaling relations \reff{beta} and
\reff{one_nu} among these exponents. Using the two other
equations we obtained the last two critical exponents $\delta$
and $\eta$. They are in agreement with those corresponding to
the Ising model on a torus.

\item
But the finite--size  scaling properties of those two systems are not
the same. In our case the position  of the maxima of
the specific heat scales near
$J_c$ with a critical exponent that does not bear any relation to the
critical behavior of the correlation length. That is, the
thermodynamic limit is achieved
faster on the tetrahedron. However, the behavior of the susceptibility
is just the same in both ones.

\end{itemize}

Our results suggest that the same analysis carried out in other types of
lattices with the same topology will yield the same conclusions: the
critical behavior does not change, although variations in their
finite--size properties are expected to hold. All of them belong to
the same universality class of the Ising model on a torus.

On the other hand, the Ising model on the  tetrahedron can be
view
as the Ising model with some non--standard boundary conditions. These
ones has the advantage that the critical behavior is reached before
than for periodic boundary conditions.

Ferdinand and Fisher \cite{fisher,barber} studied how the exponent
$\theta_{CV}$ varies for the Ising model defined on a square lattice on
a $m \times n$ torus. They concluded that $J_l(CV_{max})$ behaves as
\be
   \label{fisher_equation}
   {J_l(CV_{max}) \over J_c^{\rm torus} } = 1 + {b(\eta) \over l} +
             {\rm o}(l^{-1})
\ee
where $l = (m^{-2} + n^{-2})^{-1}$ measures the linear size of the
torus and $\eta = m/n$ its shape. They also showed that $b(\eta)$ is not
a monotonous function of $\eta$. In particular $b(\eta) > 0$ in the
range $\eta \in (\eta_0^{-1},\eta_0)$ with $\eta_0 = 3.139278$, and
$b(\eta) < 0$ in $\eta \in (0,\eta_0^{-1}) \cup (\eta_0,\infty)$.
Exactly at $\eta = \eta_0$ and $\eta_0^{-1}$ the function $b(\eta)$
vanishes\footnote{The same occurs for $\eta = 0$, $\infty$
\cite{fisher}}, so at these points the leading term in
\reff{fisher_equation} vanishes and its behavior is controlled by
the subleading term. In this model it can be written as
\be
   \left.
   {J_l(CV_{max}) \over J_c^{\rm torus} } \right|_{\eta = \eta_0}
    = 1 - {c \log l \over l^2} + \cdots
\ee
The behavior of $b(\eta)$ as a function of the shape of the torus is
explained as a highly non--trivial interplay between the different
terms which appear in the expression of the partition function. We
believe that the same feature is present in the Ising model on the
truncated tetrahedron. In this case, the shape of the lattice is fixed,
but the chosen boundary conditions are the basic ingredient which makes
the leading term in \reff{fisher_equation} vanish.

We should mention that the our conclusions may not apply to the
antiferromagnetic regime. The reason is that for such  boundary
conditions, the lattice is not bipartite. Thus, the phenomenon of
frustration may occur in that regime. This feature is absent in the
Ising model on the torus. In this case, the lattice is bipartite and
for sufficiently low temperatures we find a N\'eel ground state. This
question on the tetrahedron is currently under research.

Finally, we would like to say a few words about the continuum theory
which is attained in the thermodynamic limit. The evidence we have
found of scaling suggests a description in terms of the fields and
weights of a conformal field theory. This cannot be a trivial example
of field theory on the sphere (or on the plane), since four curvature
singularities arise which cannot be removed by conformal transformations.
Under the assumption of conformal invariance, though, we could still
stick three of the vertices together (at a point we may take as infinity)
by means of $SL(2,C)$ transformations, leaving alone a singularity
in the bulk. This picture is close in spirit to the Coulomb
gas representation of conformal field theories, but with the difference
now that not all the curvature is pinched at the point at infinity.
A conical singularity on the plane may have sensible effects on the
correlation functions of the theory. We recall here another example with
nontrivial boundary conditions, namely that of a conformal field theory
on the semiplane. In this case correlators which are taken at a finite
distance of the boundary do not measure the conformal weights of the
theory on the plane \cite{cardy}. In our model, it may not be necessary
to compute
correlators infinitely far away from the singularity to measure bulk
conformal weights, but again some dependence on the location of the points
should be expected. This point should deserve further clarification,
though its investigation in the lattice would require more powerful
computer facilities than used in the present work.

\section*{Acknowledgements}
We thank Juan Jes\'us Ruiz--Lorenzo and Miguel Angel
Mart\'{\i}n--Delgado for helpful discussions. We acknowledge the
financial support of the CICyT.

\newpage

\section*{Figure Captions}

\noindent
{\bf Figure 1:} Generic triangular block for honeycomb lattices
embeded on the tetrahedron.

\vspace{0.5cm}

\noindent
{\bf Figure 2:} Different paths through a honeycomb lattice site.

\vspace{0.5cm}

\noindent
{\bf Figure 3:} Dimer configurations around a triangle of the decorated
lattice.

\vspace{0.5cm}

\noindent
{\bf Figure 4:} Decorated lattice for the second generation. The outer
lines show the identifications of boundary links which embed the lattice
on the tetrahedron.

\vspace{0.5cm}

\noindent
{\bf Figure 5:} Values of the unsubstracted two--point correlators
$\langle \sigma_{\bf r} \sigma_{\bf 0} \rangle$ for different
separations $r$ and couplings constants $J$ = 0.57, 0.58, $\cdots$,
0.62. The least $\chi^2$ fits are also despicted.

\vspace{0.5cm}

\noindent
{\bf Figure 6:} Values of $\xi(J)^{-1}$ for those $J$ listed in Table~3.
The straight line corresponds to the $\chi^2$ fit.

\vspace{0.5cm}

\noindent
{\bf Figure 7:} Values of the position of the maxima of the specific
heat (circles) and the magnetic susceptibility (squares) with respect to
the critical coupling $J_c^{\rm torus}$. The power--law fits are also
despicted.

\vspace{0.5cm}

\noindent
{\bf Figure 8:} Power--law fits of the values of the specific
heat (squares) and the magnetic susceptibility (circles) at the
critical point $J_c$.

\vspace{0.5cm}

\noindent
{\bf Figure 9:} The same as in Figure 8 for the magnetization
$M_N(J_c)$.

\vspace{0.5cm}

\noindent
{\bf Figure 10:} Logarithmic least $\chi^2$ fit to $CV_N(J_c)$.

\newpage
%
%
\def\dd{$^\dagger$}
\def\de{$^\ddagger$}
\def\df{$^\star$}
\def\jct{$J_c^{\rm torus}$}
\section*{Table Captions}

\noindent
{\bf Table 1:}
For each lattice size $N$ we show the values of the
maximum of the specific heat and the position of such point (\cvmax,
\jcvmax) and the same for the susceptibility (\chimax, \jchimax).
Those values marked with a $\dagger$ were computed using the exact
partition function \cite{jose}, those with a $\ddagger$ using the dimer
approach (see Section~2) and those with an $\star$ by evaluating
exactly (and numerically) the
partition function. The rest were obtained by means of the Monte Carlo
simulations described in Section~3.

\vspace{0.5cm}

\noindent
{\bf Table 2:}
For each lattice size $N$ we show the values of the
specific heat \cvc, the susceptibility \chic\ and the magnetization
\mc\ evaluated at the critical point $J_c^{\rm torus}$.
The symbols posses the same
meaning as in Table~1.

\vspace{0.5cm}

\noindent
{\bf Table 3:}
Values of the inverse of the correlation length  computed
for the $N=9$ lattice and different values of the coupling $J$.

\vspace{0.5cm}

\noindent
{\bf Table 4:}
Number of Monte Carlo steps performed for each simulation
in a lattice of volume $V=12N^2$ and coupling constant $J$.

\newpage

\setcounter{table}{0}
\begin{table}[h]
\centering
\begin{tabular}{|c|c|c|c|c|} \hline
   &                &               &               &             \\
$N$&  \cvmax        & \jcvmax       & \chimax       & \jchimax    \\
   &                &               &               &             \\
\hline \hline
  &                 &               &               &             \\
1 & 1.43923551(1)\dd& 0.467332(1)\dd&1.2564278(1)\df&0.487805(1)\df \\
2 & 1.65204595(1)\dd& 0.608224(1)\dd&               &             \\
3 & 1.89238232(1)\dd& 0.634238(1)\dd& 7.91(2)       & 0.6215(3)   \\
4 & 2.07977738(1)\de& 0.643862(1)\de&               &             \\
5 & 2.23081551(1)\de& 0.648580(1)\de& 18.98(7)      & 0.6377(3)   \\
6 & 2.35682745(1)\de& 0.651275(1)\de&               &             \\
7 & 2.46477396(1)\de& 0.652973(1)\de& 34.4(1)       & 0.6444(2)   \\
9 & 2.64288048(1)\de& 0.654929(1)\de& 53.3(2)       & 0.6480(2)   \\
11& 2.786536(6)\de  & 0.65598(5)\de & 75.9(4)       & 0.6498(2)   \\
15& 3.02(1)         & 0.6571(3)     & 128.4(8)      & 0.6523(2)   \\
21& 3.30(2)         & 0.6579(3)     & 232(2)        & 0.6540(2)   \\
  &                 &               &      &    \\
\hline \end{tabular}

\vspace*{1cm}
{\bf Table 1}
\end{table}

\begin{table}[h]
\centering
\begin{tabular}{|c|c|c|c|} \hline
   &                 &                &             \\
$N$&  \cvc           & \chic          & \mc         \\
   &                 &                &             \\ \hline \hline
   &                 &                &             \\
1  & 0.99350489(1)\dd& 0.8617020(1)\df& 0.82868806(1)\df \\
2  & 1.51639951(1)\dd&                &              \\
3  & 1.81746516(1)\dd& 6.889(3)       & 0.6904(9)    \\
5  & 2.19465654(1)\de& 16.89(9)       & 0.645(1)     \\
6  & 2.32898639(1)\de&                &              \\
7  & 2.44249002(1)\de& 30.8(1)        & 0.6168(7)    \\
9  & 2.62744012(1)\de& 48.2(3)        & 0.597(1)     \\
11 & 2.77(1)         & 68.3(4)        & 0.5828(9)    \\
15 & 2.998(9)        & 116(1)         & 0.560(1)     \\
21 & 3.24(1)         & 210(2)         & 0.536(2)     \\
   &                 &                &              \\ \hline
\end{tabular}

\vspace*{1cm}
{\bf Table 2}
\end{table}

\begin{table}[h]
\centering
\begin{tabular}{|c|c|} \hline
     &         \\
$J$  & $\xi^{-1}_{N=9}$\\
     &         \\ \hline \hline
     &         \\
0.57 &  0.204(7)      \\
0.58 &  0.181(5)      \\
0.59 &  0.158(1)      \\
0.60 &  0.135(2)      \\
0.61 &  0.112(1)      \\
0.62 &  0.088(1)      \\
     &         \\ \hline
\end{tabular}

\vspace*{1cm}
{\bf Table 3}
\end{table}

\begin{table}[h]
\centering
\begin{tabular}{|r|c|r|} \hline
     &         &                 \\
$N$  &   $J$   &   MC steps      \\
     &         &                  \\ \hline
 \hline
     &         &                  \\
3    & \jct    & $3 \cdot 10^6 $  \\
3    & 0.62    & $3 \cdot 10^6 $  \\
5    & \jct    & $3 \cdot 10^6 $  \\
5    & 0.64    & $3 \cdot 10^6 $  \\
7    & \jct    & $12 \cdot 10^6$  \\
9    & \jct    & $12 \cdot 10^6$  \\
11   & \jct    & $12 \cdot 10^6$  \\
15   & \jct    & $11 \cdot 10^6$  \\
21   & \jct    & $12 \cdot 10^6$  \\ \hline
\end{tabular}

\vspace*{1cm}
{\bf Table 4}
\end{table}

\newpage
\begin{figure}[p]
\centering
\setlength{\unitlength}{1.2cm}
\begin{picture}(15,13)(2,0)

\thicklines

\multiput(5.66,2)(0,3){3}{\line(5,-3){0.83}}
\multiput(6.5,1.5)(0,3){3}{\line(5,3){0.83}}
\multiput(7.33,5)(0,3){2}{\line(5,-3){0.83}}
\put(8.16,7.5){\line(5,3){0.83}}
\put(9,5){\line(5,-3){0.83}}
\put(9.83,4.5){\line(5,3){0.83}}
\put(10.66,5){\line(5,-3){0.83}}
\put(11.5,4.5){\line(5,3){0.83}}

\multiput(5.66,3)(0,3){3}{\line(5,3){0.83}}
\multiput(6.5,3.5)(0,3){3}{\line(5,-3){0.83}}
\multiput(7.33,3)(0,3){2}{\line(5,3){0.83}}
\put(8.16,3.5){\line(5,-3){0.83}}

\put(9,6){\line(5,3){0.83}}
\put(9.83,6.5){\line(5,-3){0.83}}
\put(10.66,6){\line(5,3){0.83}}
\put(11.5,6.5){\line(5,-3){0.83}}
\multiput(6.5,3.5)(0,3){2}{\line(0,1){1}}
\multiput(8.16,3.5)(0,3){2}{\line(0,1){1}}
\multiput(9.83,3.5)(0,3){2}{\line(0,1){1}}

\multiput(5.66,2)(0,6){2}{\line(0,1){1}}
\multiput(7.33,2)(0,6){2}{\line(0,1){1}}
\put(9,5){\line(0,1){1}}
\put(10.66,5){\line(0,1){1}}
\put(12.33,5){\line(0,1){1}}
\multiput(5.66,1)(0,1){9}{\line(0,1){0.4}}
\multiput(5.66,1.6)(0,1){9}{\line(0,1){0.4}}

\multiput(13.16,5.5)(-0.833,-0.5){9}{\line(-5,-3){0.333}}
\multiput(12.66,5.2)(-0.833,-0.5){9}{\line(-5,-3){0.333}}
\multiput(13.16,5.5)(-0.833,0.5){9}{\line(-5,3){0.333}}
\multiput(12.66,5.8)(-0.833,0.5){9}{\line(-5,3){0.333}}
\put(7.33,5){\line(0,1){0.4}}
\put(7.33,6){\line(0,-1){0.4}}
\put(8.16,4.5){\line(5,3){0.333}}
\put(9,5){\line(-5,-3){0.333}}
\put(8.16,6.5){\line(5,-3){0.333}}
\put(9,6){\line(-5,3){0.333}}

\end{picture}
\caption{}
\end{figure}

\newpage
\begin{figure}[p]
\centering
\setlength{\unitlength}{1.2cm}
\begin{picture}(15,13)(2,0)

\put(6.5,3.5){\line(5,3){0.83}}
\put(7.33,4){\line(5,-3){0.83}}
\put(7.33,4){\line(0,1){1}}
\put(6.417,3.55){\line(5,3){0.83}}
\put(7.247,4.05){\line(0,1){1}}
\put(7.13,2.5){\mbox{\Large (c)}}

\put(6.5,8.5){\line(5,3){0.83}}
\put(7.33,9){\line(5,-3){0.83}}
\put(7.33,9){\line(0,1){1}}
\put(7.13,7.5){\mbox{\Large (a)}}

\put(9.83,3.5){\line(5,3){0.83}}
\put(10.66,4){\line(5,-3){0.83}}
\put(10.66,4){\line(0,1){1}}
\put(9.83,3.403){\line(5,3){0.83}}
\put(10.66,3.903){\line(5,-3){0.83}}
\put(10.46,2.5){\mbox{\Large (d)}}

\put(9.83,8.5){\line(5,3){0.83}}
\put(10.66,9){\line(5,-3){0.83}}
\put(10.66,9){\line(0,1){1}}
\put(10.743,9.05){\line(5,-3){0.83}}
\put(10.743,9.05){\line(0,1){1}}
\put(10.46,7.5){\mbox{\Large (b)}}

\end{picture}
\caption{}
\end{figure}

\newpage
\begin{figure}[p]
\centering
\setlength{\unitlength}{1.2cm}
\begin{picture}(15,13)(2,0)

\newsavebox{\toya}
\savebox{\toya}(2,3)[bl]{\begin{picture}(2,3)(-2,0)

\put(-0.05,0.7222){\line(0,1){0.8334}}
\put(0.05,0.7222){\line(0,1){0.8334}}
\put(0.2083,0.375){\line(-3,5){0.2083}}
\put(-0.2083,0.375){\line(3,5){0.2083}}
\put(-0.2083,0.375){\line(1,0){0.4167}}
\put(0.2333,0.4165){\line(5,-3){0.62505}}
\put(0.1883,0.3335){\line(5,-3){0.62505}}
\put(-0.2333,0.4165){\line(-5,-3){0.62505}}
\put(-0.1833,0.3335){\line(-5,-3){0.62505}}
\put(-0.2,-1.125){\mbox{\Large (a)}}
\end{picture}}

\newsavebox{\toyb}
\savebox{\toyb}(2,3)[bl]{\begin{picture}(2,3)(-2,0)

\put(0.0,0.7222){\line(0,1){0.8334}}
\put(0.2083,0.375){\line(-3,5){0.2083}}
\put(0.2913,0.425){\line(-3,5){0.2083}}

\put(-0.2083,0.375){\line(3,5){0.2083}}
\put(-0.2083,0.375){\line(1,0){0.4167}}
\put(0.2083,0.375){\line(5,-3){0.62505}}

\put(-0.2333,0.4165){\line(-5,-3){0.62505}}
\put(-0.1833,0.3335){\line(-5,-3){0.62505}}
\put(-0.2,-1.125){\mbox{\Large (b)}}

\end{picture}}

\newsavebox{\toyc}
\savebox{\toyc}(2,3)[bl]{\begin{picture}(2,3)(-2,0)

\put(0.0,0.7222){\line(0,1){0.8334}}
\put(0.2083,0.375){\line(-3,5){0.2083}}

\put(-0.2083,0.375){\line(3,5){0.2083}}
\put(-0.2913,0.425){\line(3,5){0.2083}}

\put(-0.2083,0.375){\line(1,0){0.4167}}

\put(0.2333,0.4165){\line(5,-3){0.62505}}
\put(0.1833,0.3335){\line(5,-3){0.62505}}

\put(-0.2083,0.375){\line(-5,-3){0.62505}}
\put(-0.2,-1.125){\mbox{\Large (c)}}

\end{picture}}

\newsavebox{\toyd}
\savebox{\toyd}(2,3)[bl]{\begin{picture}(2,3)(-2,0)

\put(-0.05,0.7222){\line(0,1){0.8334}}
\put(0.05,0.7222){\line(0,1){0.8334}}

\put(0.2083,0.375){\line(-3,5){0.2083}}
\put(-0.2083,0.375){\line(3,5){0.2083}}

\put(-0.2083,0.375){\line(1,0){0.4167}}
\put(-0.2083,0.275){\line(1,0){0.4167}}

\put(0.2083,0.375){\line(5,-3){0.62505}}
\put(-0.2083,0.375){\line(-5,-3){0.62505}}
\put(-0.2,-1.125){\mbox{\Large (d)}}

\end{picture}}

\put(4.7,3.5){\usebox{\toyc}}
\put(4.7,8.5){\usebox{\toya}}
\put(8.83,3.5){\usebox{\toyd}}
\put(8.83,8.5){\usebox{\toyb}}

\end{picture}
\caption{}
\end{figure}

\newpage
\begin{figure}[p]
\centering
\setlength{\unitlength}{1.8cm}
\begin{picture}(5,9.5)(-0.3,-1)

\thinlines
\put(5.9,5){\oval(0.4,1.6667)[r]}
\put(5.9,5){\oval(0.8,5)[r]}
\put(5.9,5){\oval(1.2,8.333)[r]}
\put(-0.9,5){\oval(0.4,1.6667)[l]}
\put(-0.9,5){\oval(0.8,5)[l]}
\put(-0.9,5){\oval(1.2,8.333)[l]}
\put(2.5,-0.2){\oval(3,0.4)[b]}
\put(2.5,10.2){\oval(3,0.4)[t]}
\multiput(-0.5,5)(0.48,0){13}{\line(1,0){0.24}}
\multiput(-0.5,5)(0.24,0.4){13}{\line(3,5){0.12}}
\multiput(-0.5,5)(0.24,-0.4){13}{\line(3,-5){0.12}}
\multiput(2.5,0)(0.24,0.4){13}{\line(3,5){0.12}}
\multiput(2.5,10)(0.24,-0.4){13}{\line(3,-5){0.12}}

\thicklines
\newsavebox{\arrr}
\savebox{\arrr}(1,1)[bl]{\begin{picture}(1,1)
\put(0,0){\line(-5,3){0.2}}
\put(0,0){\line(-5,-3){0.2}}
\end{picture}}
\newsavebox{\arrl}
\savebox{\arrl}(1,1)[bl]{\begin{picture}(1,1)
\put(0,0){\line(5,3){0.2}}
\put(0,0){\line(5,-3){0.2}}
\end{picture}}

\newsavebox{\arrul}
\savebox{\arrul}(1,1)[bl]{\begin{picture}(1,1)
\put(0,0){\line(5,-3){0.2}}
\put(0,0){\line(0,-1){0.2332}}
\end{picture}}
\newsavebox{\arrdl}
\savebox{\arrdl}(1,1)[bl]{\begin{picture}(1,1)
\put(0,0){\line(5,3){0.2}}
\put(0,0){\line(0,1){0.2332}}
\end{picture}}

\newsavebox{\arrdr}
\savebox{\arrdr}(1,1)[bl]{\begin{picture}(1,1)
\put(0,0){\line(-5,3){0.2}}
\put(0,0){\line(0,1){0.2332}}
\end{picture}}

\newsavebox{\toy}
\savebox{\toy}(3,2)[bl]{\begin{picture}(3,2)
\put(1.625,-0.2083){\line(0,1){0.4167}}
\put(1.278,0.0){\line(5,3){0.3472}}
\put(1.278,0.0){\line(5,-3){0.3472}}
\put(0.7222,0.0){\line(1,0){0.5556}}
\put(0.375,0.2083){\line(5,-3){0.3472}}
\put(0.375,-0.2083){\line(5,3){0.3472}}
\put(0.375,-0.2083){\line(0,1){0.4167}}
\put(0.375,0.2083){\line(-3,5){0.25}}
\put(1.625,0.2083){\line(3,5){0.25}}
\end{picture}}

\newsavebox{\cort}
\savebox{\cort}(3,2)[bl]{\begin{picture}(3,2)
\put(1.625,0.0){\line(0,1){0.20835}}
\put(1.278,0.0){\line(5,3){0.3472}}
\put(0.7222,0.0){\line(1,0){0.5556}}
\put(0.375,0.2083){\line(5,-3){0.3472}}
\put(0.375,0.0){\line(0,1){0.20835}}
\put(0.375,0.2083){\line(-3,5){0.25}}
\put(1.625,0.2083){\line(3,5){0.25}}
\end{picture}}

\newsavebox{\two}
\savebox{\two}(3,2)[bl]{\begin{picture}(3,2)
\put(1.625,-0.2083){\line(0,1){0.20835}}
\put(1.278,0.0){\line(5,-3){0.3472}}
\put(0.7222,0.0){\line(1,0){0.5556}}
\put(0.375,-0.2083){\line(5,3){0.3472}}
\put(0.375,-0.2083){\line(0,1){0.20835}}
\end{picture}}
\newsavebox{\arr}
\savebox{\arr}(3,2)[bl]{\begin{picture}(3,2)
\put(-0.125,0.2083){\line(5,-3){0.3472}}
\put(-0.125,-0.2083){\line(5,3){0.3472}}
\put(-0.125,-0.2083){\line(0,1){0.4167}}
\put(-0.125,0.2083){\line(-3,5){0.25}}
\put(0.2222,0.0){\line(1,0){0.2778}}
\end{picture}}

\newsavebox{\abj}
\savebox{\abj}(3,2)[bl]{\begin{picture}(3,2)
\put(0.125,-0.2083){\line(0,1){0.4167}}
\put(-0.222,0.0){\line(5,3){0.3472}}
\put(-0.222,0.0){\line(5,-3){0.3472}}
\put(0.125,0.2083){\line(3,5){0.25}}
\put(-0.5,0.0){\line(1,0){0.2778}}
\end{picture}}
\put(0,0){\usebox{\cort}}
\multiput(0,1.6667)(0,1.6667){5}{\usebox{\toy}}
\multiput(1.5,0.8333)(0,1.6667){6}{\usebox{\toy}}
\put(3.0,0.0){\usebox{\cort}}
\multiput(3.0,1.6667)(0,1.6667){5}{\usebox{\toy}}
\multiput(0.0,0.8333)(0,1.6667){6}{\usebox{\abj}}
\multiput(5.0,0.8333)(0,1.6667){6}{\usebox{\arr}}
\put(0.0,10.0){\usebox{\two}}
\put(3.0,10.0){\usebox{\two}}
\multiput(1.1,1.666667)(0,1.666667){5}{\usebox{\arrr}}
\multiput(2.6,0.83333)(0,1.666667){6}{\usebox{\arrr}}
\multiput(4.1,0.0)(0,1.666667){7}{\usebox{\arrr}}
\multiput(-0.4,0.83333)(0,1.666667){3}{\usebox{\arrr}}
\multiput(0.9,0.0)(0,10.0){2}{\usebox{\arrl}}
\multiput(5.4,5.83333)(0,1.666667){3}{\usebox{\arrl}}
\multiput(0.2,2.1666667)(0,1.666667){5}{\usebox{\arrul}}
\multiput(1.7,1.33333)(0,1.666667){5}{\usebox{\arrul}}
\multiput(3.2,0.5)(0,1.666667){6}{\usebox{\arrul}}
\multiput(4.7,1.33333)(0,1.666667){6}{\usebox{\arrul}}
\multiput(0.2,1.1666667)(0,1.666667){6}{\usebox{\arrdl}}
\multiput(1.7,0.33333)(0,1.666667){6}{\usebox{\arrdl}}
\multiput(3.2,1.1666667)(0,1.666667){6}{\usebox{\arrdl}}
\multiput(4.7,0.33333)(0,1.666667){6}{\usebox{\arrdl}}
\put(0.3,0.333333){\usebox{\arrdr}}
\put(1.8,9.5){\usebox{\arrdr}}
\end{picture}
\caption{}
\end{figure}

\end{document}